\documentclass{article}
\usepackage{spconf,amsmath,graphicx}
\usepackage[ruled,vlined]{algorithm2e}

\begin{document}
\ninept

\title{PixelHop++: A Small Successive-Subspace-Learning-Based \\
(SSL-based) Model for Image Classification}

\name{Yueru Chen$^{*}$, Mozhdeh Rouhsedaghat$^{*}$, Suya You$^{+}$, 
Raghuveer Rao$^{+}$ and C.-C. Jay Kuo$^{*}$}
\address{$^{*}$ University of Southern California, Los Angeles,
California, USA \\ $^{+}$ Army Research Laboratory, Adelphi, Maryland, USA}

\maketitle
\begin{abstract}

The successive subspace learning (SSL) principle was developed and used
to design an interpretable learning model, known as the PixelHop method,
for image classification in our prior work. Here, we propose an improved
PixelHop method and call it PixelHop++. First, to make the PixelHop
model size smaller, we decouple a joint spatial-spectral input tensor to
multiple spatial tensors (one for each spectral component) under the
spatial-spectral separability assumption and perform the Saab transform
in a channel-wise manner, called the channel-wise (c/w) Saab transform.
Second, by performing this operation from one hop to another
successively, we construct a channel-decomposed feature tree whose leaf
nodes contain features of one dimension (1D). Third, these 1D features
are ranked according to their cross-entropy values, which allows us to
select a subset of discriminant features for image classification. In
PixelHop++, one can control the learning model size of fine-granularity,
offering a flexible tradeoff between the model size and the
classification performance. We demonstrate the flexibility of
PixelHop++ on MNIST, Fashion MNIST, and CIFAR-10 three datasets. 

\end{abstract}

\begin{keywords}
Small learning models, interpretable learning models,
successive subspace learning, image classification.
\end{keywords}

\section{Introduction}\label{sec:intro}

The design of small machine learning models is a hot research topic in
recent years since small models are essential to mobile and edge
computing applications.  There has been a lot of research dedicated to
neural network model compression and acceleration, {\em e.g.}
\cite{cheng2018model, frankle2018lottery,han2015deep,
hubara2016binarized, rastegari2016xnor}.  Techniques such as parameter
pruning, quantization, binarization and sharing, low-rank factorization,
transferred filters, knowledge distillation, etc. have been applied to
larger network models to achieve this goal. Another path is to design
small network models from the scratch. Examples include SqueezeNet
\cite{iandola2016squeezenet}, SquishedNets
\cite{shafiee2017squishednets} and SqueezeNet-DSC
\cite{santos2018reducing}. Being similar to the situation with large
neural-network-based learning models, the underlying mechanism of small
learning models remains to be a mystery. 

Being inspired by study on deep learning networks, the successive
subspace learning (SSL) principle was recently proposed. It has been
used to design two interpretable machine learning models -- PixelHop for
image classification \cite{chen2020pixelhop} and PointHop for point
cloud classification \cite{zhang2019pointhop}.  Since no backpropagation
is needed in SSL-based model training, the training can be done
efficiently.  In this work\footnote{This research was supported by the
U.S. Army Research Laboratory's External Collaboration Initiative (ECI)
of the Director's Research Initiative (DRIA) program. The views and
conclusions contained in this document are those of the authors and
should not be interpreted as representing the official policies, either
expressed or implied, of the U.S. Army Research Laboratory or the U.S.
Government. The U.S.  Governments are authorized to reproduce and
distribute reprints for Government purposes notwithstanding any
copyright notation hereon.}, we focus on the model size (in terms of
model parameters) of PixelHop and propose several ideas to reduce its
model size, resulting in a new method called PixelHop++. 

This work has several contributions. First, we point out the weak
correlation of different spectral components of the Saab transform,
which is used in PixelHop for dimension reduction. Then, we exploit this
property to design a channel-wise (c/w) Saab transform, which can reduce
the filter size as well as the memory requirement for filter computation
in PixelHop++.  Second, we propose a novel tree-decomposed feature
representation method whose leaf node provides a scalar (or 1D) feature.
By concatenating leaf node's features, we obtain a feature vector of
higher dimension for PixelHop++.  Third, we compute the cross-entropy
value of each feature and order them from the lowest to highest. The
feature of lower cross-entropy has higher discriminant power. As a
result, we can find a proper subset of features that are suitable for
the classification task.  In PixelHop++, one can control the learning
model size of fine-granularity, offering a flexible tradeoff between the
model size and the classification performance.  We demonstrate the
flexibility of PixelHop++ on MNIST, Fashion MNIST, and CIFAR-10 three
datasets. 

\section{Background Review}\label{sec:review} 
\begin{figure*}[!ht]
\centering
\includegraphics[width=0.85\linewidth]{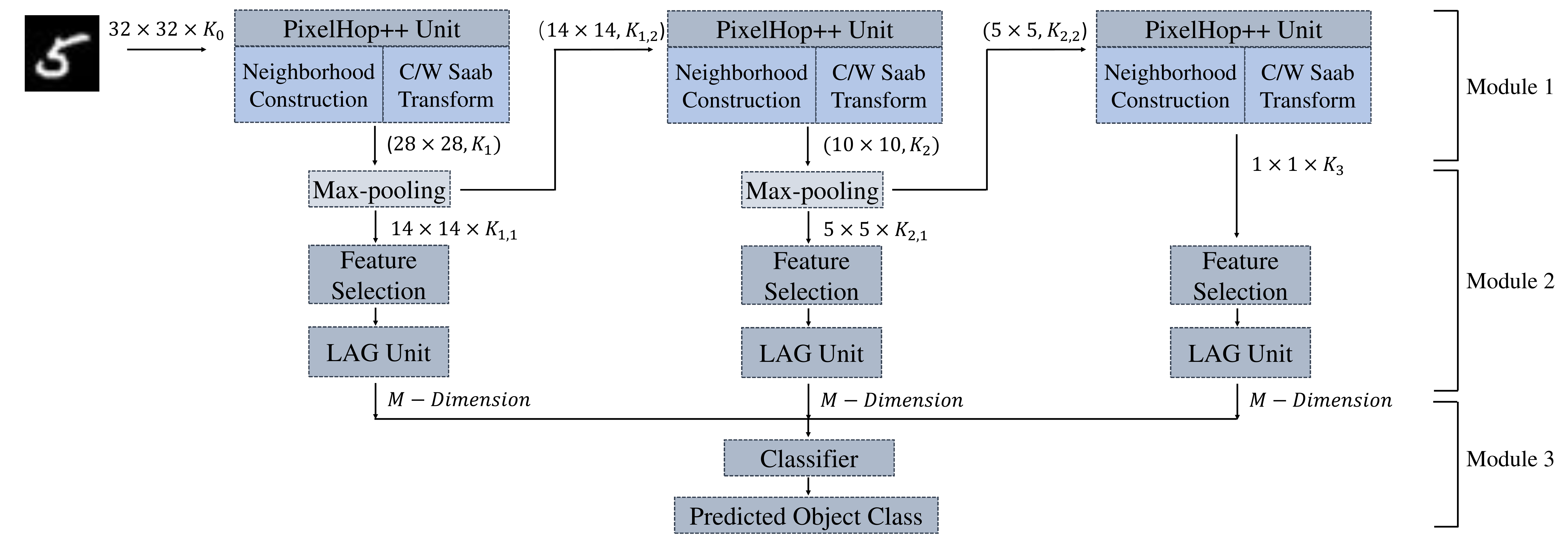}
\caption{The block diagram of the PixelHop++ method that contains
three PixelHop++ Units in cascade.}\label{fig:overview}
\end{figure*}
Subspace learning is one of the fundamental problems in signal/image
processing and computer vision \cite{turk1991eigenfaces,
bouwmans2009subspace, kriegel2009clustering, lu2011survey, gu2011joint,
wang2015joint}. The subspace method learns a projection from an input
data space into a lower-dimensional subspace which serves as an
approximation to the input space.  Being inspired by the deep learning
(DL) framework and built upon the foundation in \cite{kuo2017cnn,
kuo2018data, kuo2016understanding, kuo2019interpretable}, Kuo {\em et
al.} proposed the successive subspace learning (SSL) principle to design
interpretable machine learning models. Concrete examples include
PixelHop \cite{chen2020pixelhop} and PointHop \cite{zhang2019pointhop},
which are designed for image and point cloud classification problems,
respectively.  Their model parameters are determined stage-by-stage in a
feedforward manner without any backpropagation (BP). 

The PixelHop method contains three main modules: 1) successive
near-to-far neighborhood expansion and unsupervised dimension reduction;
2) supervised dimension reduction via designed label-assisted regression
(LAG); 3) feature concatenation and decision making.  PixelHop extracts
features successively from a pixel and its near-, mid- and far-range
neighborhoods in multiple stages, where each stage corresponds to one
PixelHop unit. To control the rapid growth of the output dimension of a
PixelHop unit, the Saab (subspace approximation with adjusted bias)
transform \cite{kuo2019interpretable} is adopted for unsupervised
dimension reduction.

\section{Proposed PixelHop++ Method}\label{sec:method} 

The block diagram of the PixelHop++ method for tiny images (size $32
\times 32$) is shown in Fig. \ref{fig:overview}.  It has three
PixelHop++ units in cascade. In all of them, the Saab transform kernels
are of the same spatial dimension ($5 \times 5$). Also, We apply the max
pooling operation to filtered outputs in the first and the second
PixelHop units. As compared with PixelHop, PixelHop++ has the following
three modifications. 
\begin{enumerate}
\item We replace the traditional Saab transform with the channel-wise
(c/w) Saab transform.
\item A novel tree-decomposed feature representation is constructed.
\item We order leaf node's features based on their cross-entropy 
values and use them to select a feature subset. 
\end{enumerate}
They will be detailed below.

\subsection{Channel-wise (c/w) Saab Transform}\label{subsec:cwSaab}

Module 1 in Fig. \ref{fig:overview} is an unsupervised feature learning
module.  By arguing that the Saab transform can be approximated by the
tensor product of two separable transforms - one along the spatial
domain and the other along the spectral domain, we can reduce the size
of the Saab transform.  The Saab transform is a variant of the PCA
(Principal Component Analysis) transform. Since PCA can decorrelate the
covariance matrix into a diagonal matrix, all channel components are
decoupled.  Although the Saab transform is not identical with the PCA
transform, we expect Saab coefficients to be weakly correlated in the
spectral domain. 

To validate the spatial-spectral separability assumption, we show the
average correlations of Saab coefficients at the outputs of the first,
the second and the third PixelHop++ units in Table \ref{table:corr}.
The first two rows in the table indicate the averaged spatial
correlation of a window of size $5 \times 5$ at the output of the first
(Spatial 1) and the second PixelHop++ units (Spatial 2) for a fixed
spectral component.  The last three rows indicate the averaged spectral
correlation at the output of the first (Spectral 1), the second
(Spectral 2) and the third (Spectral 3) Pixelhop++ units at the center
pixel location. Only outputs of AC filters are used in the computation.
We see that spectral correlations are weaker than spatial correlations.
This is especially obvious for the CIFAR-10 dataset.  Furthermore, these
correlations are weaker as we go into deeper PixelHop++ units. 

\begin{table}[h!]
\centering
\scriptsize
\caption{Averaged correlations of filtered AC outputs from the first to 
the third Pixelhop units with respect to the MNIST, Fashion MNIST and 
CIFAR-10 datasets.}\label{table:corr}
\begin{tabular}{cccc} \hline
 Dataset    & MNIST  & Fashion MNIST & CIFAR-10 \\ \hline
 Spatial 1   & $0.48 \pm 0.05$ & $0.51 \pm 0.03$ & $0.53 \pm 0.03$ \\ \hline
 Spatial 2  & $0.22 \pm 0.03$ & $0.29 \pm 0.05$ & $0.27 \pm 0.06$ \\ \hline
 Spectral 1   & $0.33 \pm 0.07$ & $0.12 \pm 0.02$ & $0.0156 \pm 0.0005$ \\ \hline
 Spectral 2  &$0.18 \pm 0.02$ & $0.13 \pm 0.01$ & $0.0188 \pm 0.0004$ \\ \hline
 Spectral 3  &$0.0099 \pm 0.0001$ & $0.0082 \pm 0.0001$ & $0.0079 \pm 0.0004$ \\ \hline
\end{tabular}
\end{table}

The weak spectral correlation of Saab coefficients allows us to
approximately decompose the joint spatial-spectral input tensor of
dimension $5\times5 \times K_i$, $i=1,2$, to the $(i+1)$th PixelHop++
Unit into $K_i$ spatial tensors of size $5\times5$ (i.e., one for each
spectral component), respectively. Then, instead of performing the
traditional Saab transform of high dimension in PixelHop, we apply $K_i$
channel-wise (c/w) Saab transforms to each of the spatial tensors in
PixelHop++. 

\begin{figure}[ht!]
\centering
\includegraphics[width=0.9\linewidth]{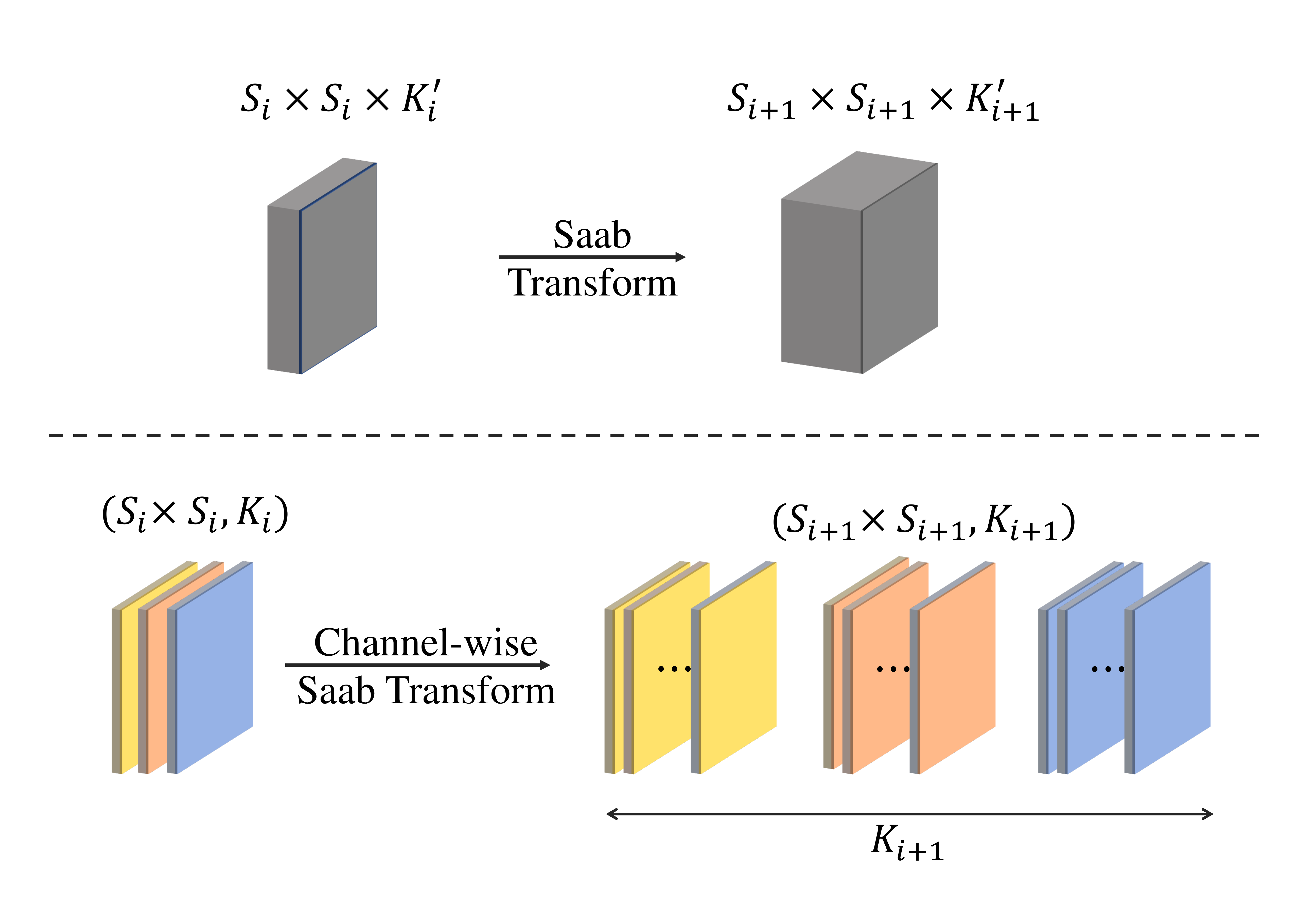}
\caption{Comparison of the traditional Saab transform and the proposed 
c/w Saab transform.}\label{fig:channel_wise}
\end{figure}

The traditional Saab transform and the c/w Saab transform are compared
in Fig.  \ref{fig:channel_wise} from the whole image viewpoint. The
traditional Saab transform takes an input image of dimension $S_{i}
\times S_{i} \times K'_{i}$ and generates an output image of dimension
$S_{i+1} \times S_{i+1} \times K'_{i+1}$ after the max-pooling operation
that pools from a grid of size $S_{i} \times S_{i}$ to a grid of size
$S_{i+1} \times S_{i+1}$. The c/w Saab transform takes $K_i$
channel-images of dimension $S_{i} \times S_{i}$ as the input and
generates $K_{i+1}$ output images of dimension $S_{i+1} \times S_{i+1}$
after max-pooling. 

We adopt the same kernel and bias design principle of the Saab transform
in the design of the c/w Saab transform. Their main difference lies in
the input tensor size. The dimension of input tensors for PixelHop++ is
the spatial neighborhood size ($5\times5=25$ in the current example)
while that for PixelHop is the multiplication of the spectral dimension
and the spatial dimension ($25 \times K'$). Separable transforms, in
general, facilitates computation and allows a smaller set of kernel
parameters. 

\subsection{Tree-decomposed Feature Representation}\label{sec:tree-decomposition}

\begin{figure}[ht!]
\centering
\includegraphics[width=0.98\linewidth]{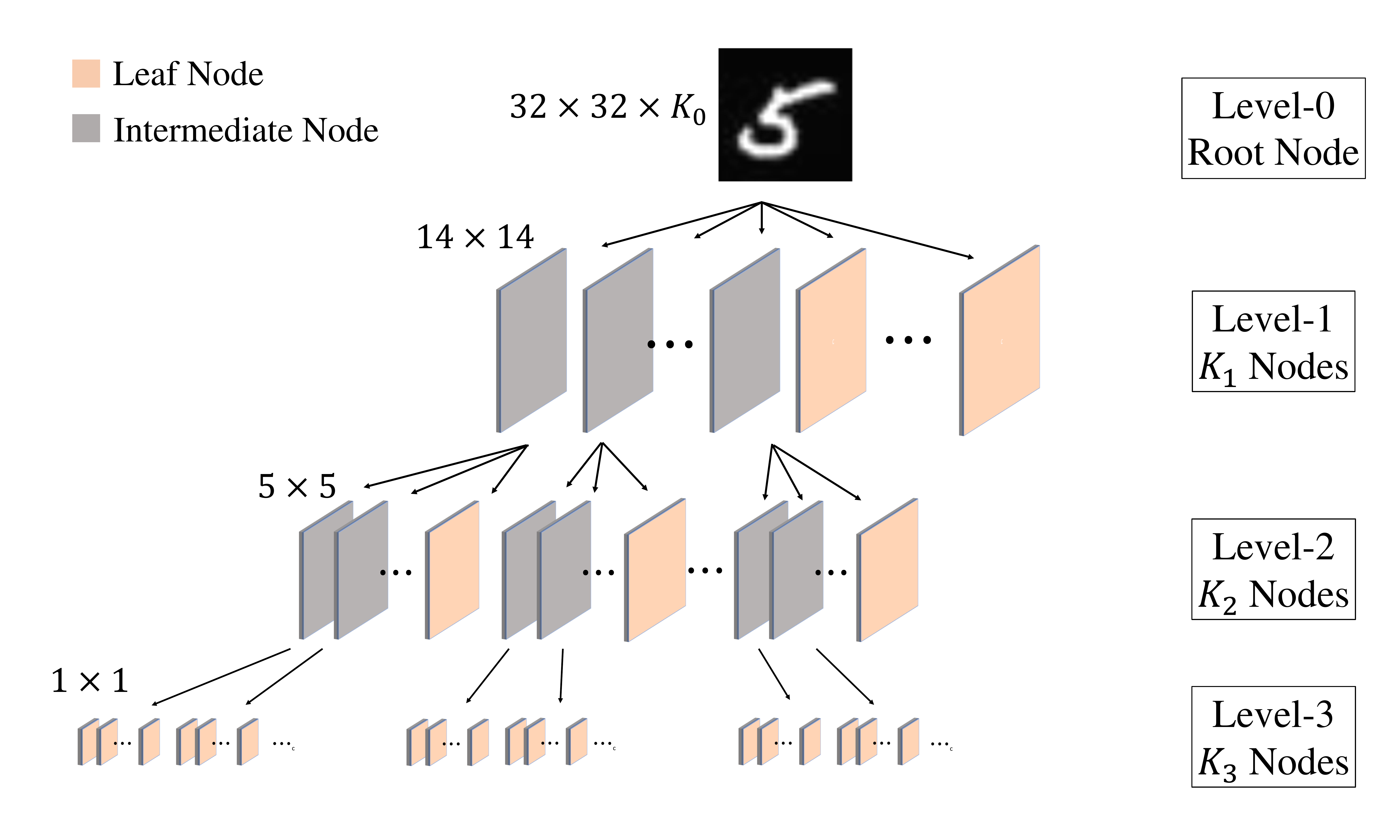}
\caption{Illustration of the tree-decomposed feature representation.}
\label{fig:tree_representation}
\end{figure}

In this subsection, we propose a new feature representation method,
called the tree-decomposed feature representation as illustrated in Fig.
\ref{fig:tree_representation}.  The root node of the tree is the input
image of dimension $32 \times 32 \times K_{0}$, where $K_0=1$ and $3$
for the gray-scale and color images, respectively.  We normalize the
total energy of the root node to unity. The first PixelHop++ unit yields
the first level child nodes.  This unit applies the c/w Saab transform
of $5 \times 5$ transform kernels to input images with stride equal to
one. Its output contains multiple response maps of dimension $28 \times
28$, where the boundary effect is taken into account. We apply the
standard $(2\times2)$-to-$(1\times1)$ max-pooling to reduce the spatial
redundancy of the response maps.  The final output of first PixelHop++
unit is a set of $K_{1}$ response maps of dimension $14 \times 14$.
Each response map is a child node of the root node. The energy of a
child node is the multiplication of its parent node energy and its
normalized energy with respect to its parent node. If the energy of a
child node is smaller than the pre-set threshold $T$, we treat it as a
leaf node and the total energy of the response map is used as the
feature of the node. If the energy of a child node is larger than
threshold $T$, its response map will be used as the input to the next
stage for further processing. It is called an intermediate node.  For
example, there are $K_{i}$ nodes at the output of the $i$th PixelHop++
unit, $i=1,2$. We use $K_{i,1}$ and $K_{i,2}$ to denote the number of
leaf and intermediate nodes at the $i$th level in Fig.
\ref{fig:overview}. Clearly, $K_{i,1}+K_{i,2}=K_i$. 

We conduct the same operation in each PixelHop++ unit successively until
the last stage is met. Then, we obtain a tree-decomposed feature
representation as shown in Fig. \ref{fig:tree_representation}, whose
leaf node has an associated feature that corresponds to the energy of
the response map of one spectral component.  The spectral components at
different tree levels have different receptive field sizes. The
associated features are useful for the image classification task. 
 
\subsection{Cross-Entropy-Guided Feature Selection}\label{sec:feature}

The tree-decomposed feature representation process is unsupervised. It
provides task-independent features. Next, we need to find a link from
features to desired labels. In this subsection, we will develop new
Module 2 in Fig. \ref{fig:overview} based on this representation. 
First, we compute the cross-entropy value for each feature at the leaf
node via 
\begin{equation}
L =\sum_{j=1}^{J}L_j, \quad L_j =-\sum_{c=1}^{M}y_{j,c}log(p_{j,c}), 
\end{equation}
where $M$ is the class number, $y_{j,c}$ is binary indicator to show
whether sample $j$ is correctly classified, and $p_{j,c}$ is the
probability that sample $j$ belongs to class $c$.  The lower the
cross-entropy, the higher the discriminant power. We order features from
the smallest to the largest cross-entropy scores and select the top
$N_S$ features. This new cross-entropy-guided feature selection process
can reduce the model size of the label-assisted regression (LAG) unit. 

Finally, we concatenate $M$ features from each PixelHop++ unit to form a
feature vector and fed it into a classifier in Module 3, where a simple
linear least-squared regressor is adopted in our experiment.  There are
two hyperparameters, $T$ and $N_S$, which can be used to control the
model size flexibly. We will study their impacts on classification
accuracy in Sec.  \ref{subsec:hyper}. 

\section{Experiments}\label{sec:experiment} 

\begin{figure*}[!ht]
\begin{minipage}[b]{0.31\linewidth}
\centering
\centerline{\includegraphics[width=0.97\linewidth]{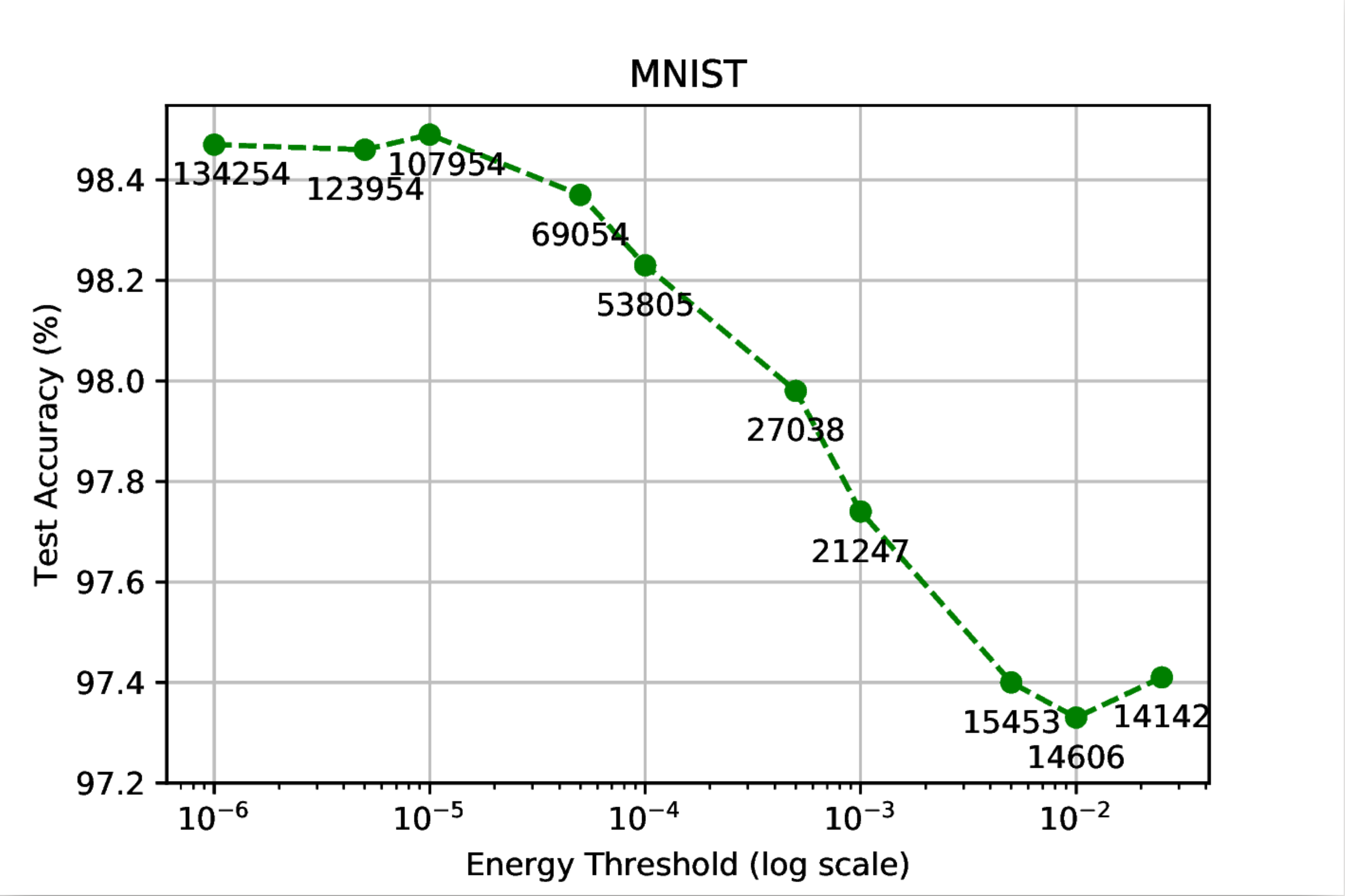}}
\end{minipage} 
\hfill
\begin{minipage}[b]{0.31\linewidth}
\centering
\centerline{\includegraphics[width=0.97\linewidth]{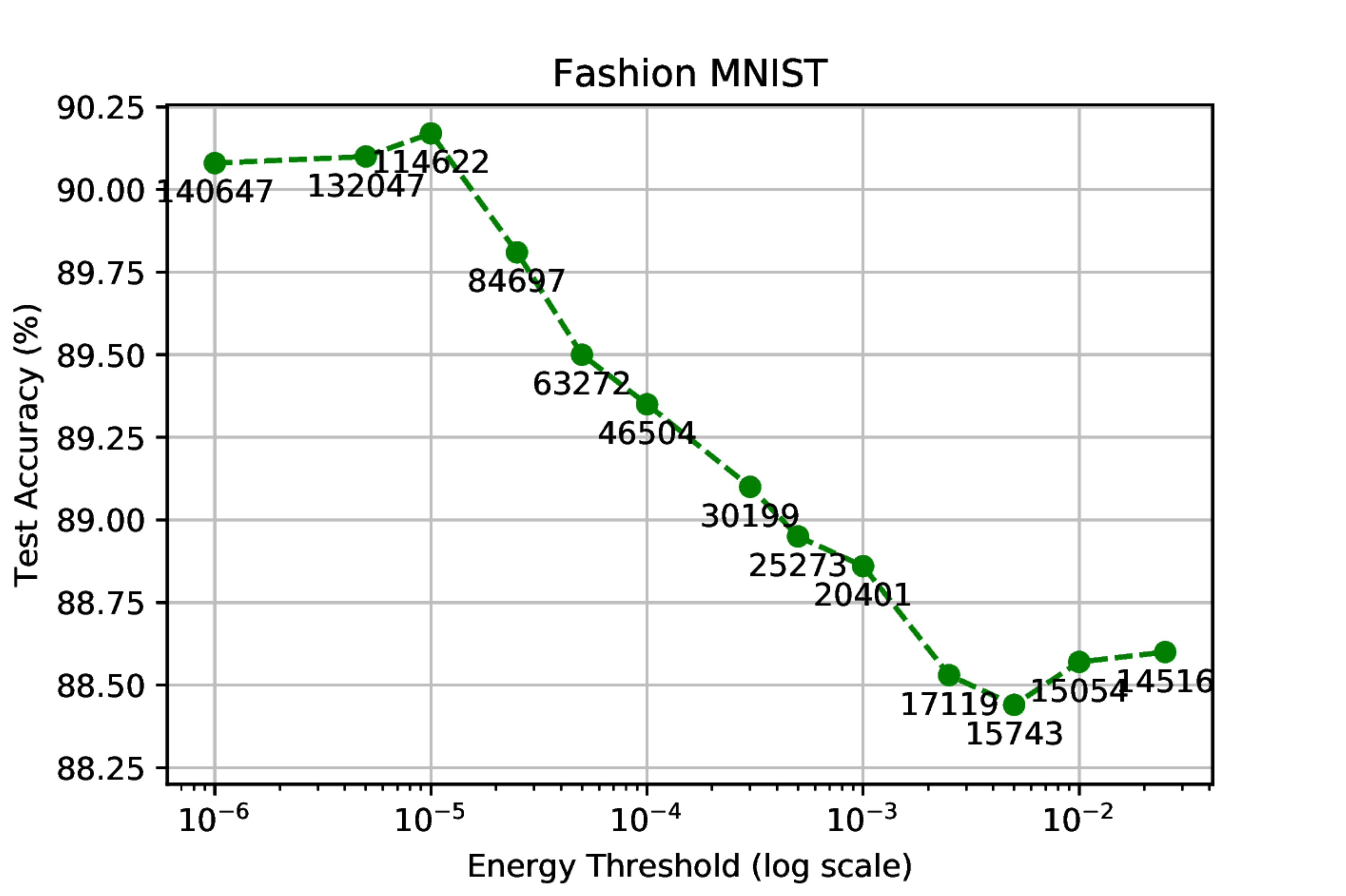}}
\end{minipage}
\hfill
\begin{minipage}[b]{0.31\linewidth}
\centering
\centerline{\includegraphics[width=0.97\linewidth]{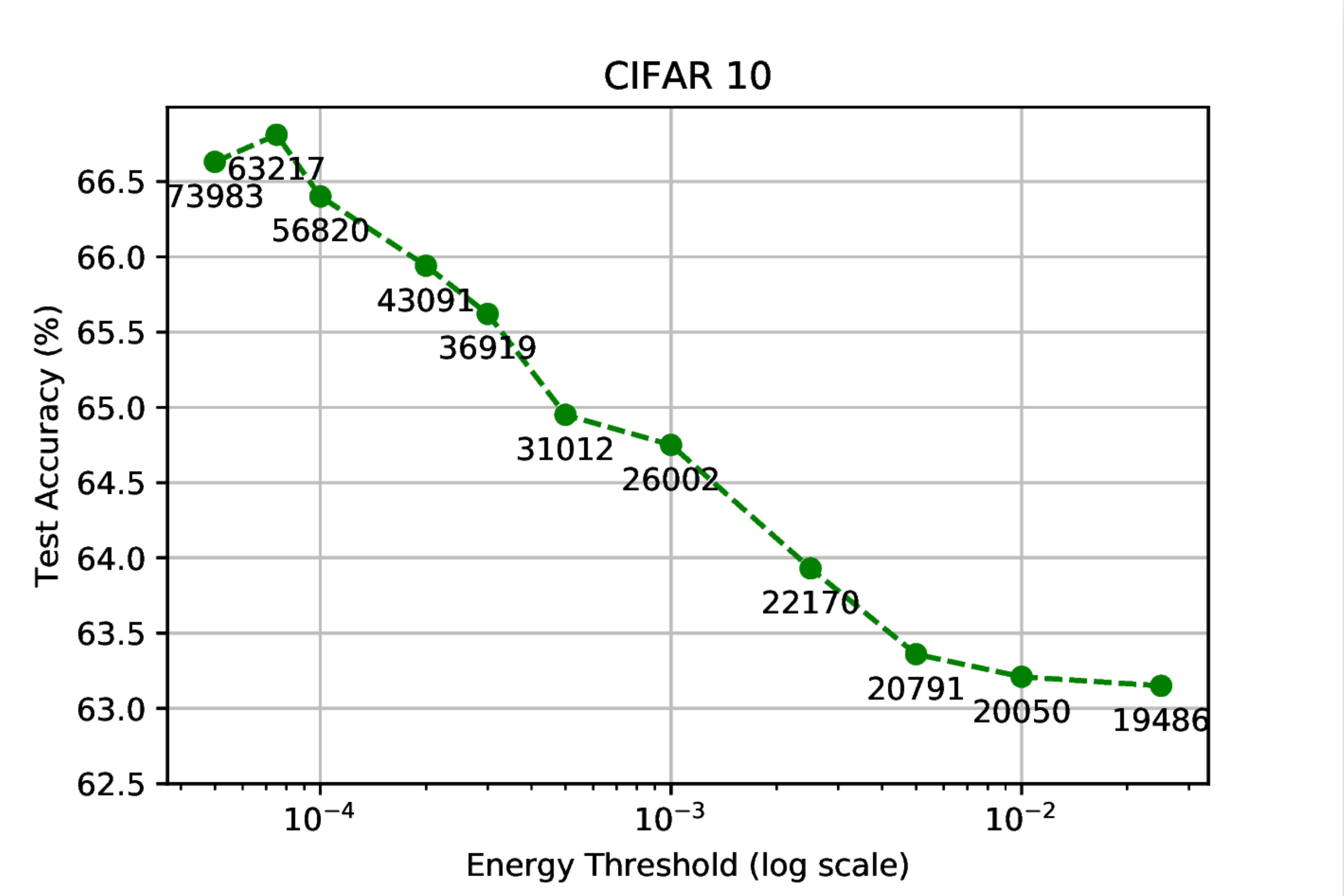}}
\end{minipage}
\caption{The relation between the test accuracy (\%) and energy threshold
$T$ in PixelHop++ for MNIST, Fashion MNIST and CIFAR-10, where the
number of model parameters in Module 1 is shown at each operational
point.} \label{fig:energy}
\end{figure*}


\begin{figure*}[!ht]
\begin{minipage}[b]{0.31\linewidth}
\centering
\centerline{\includegraphics[width=0.97\linewidth]{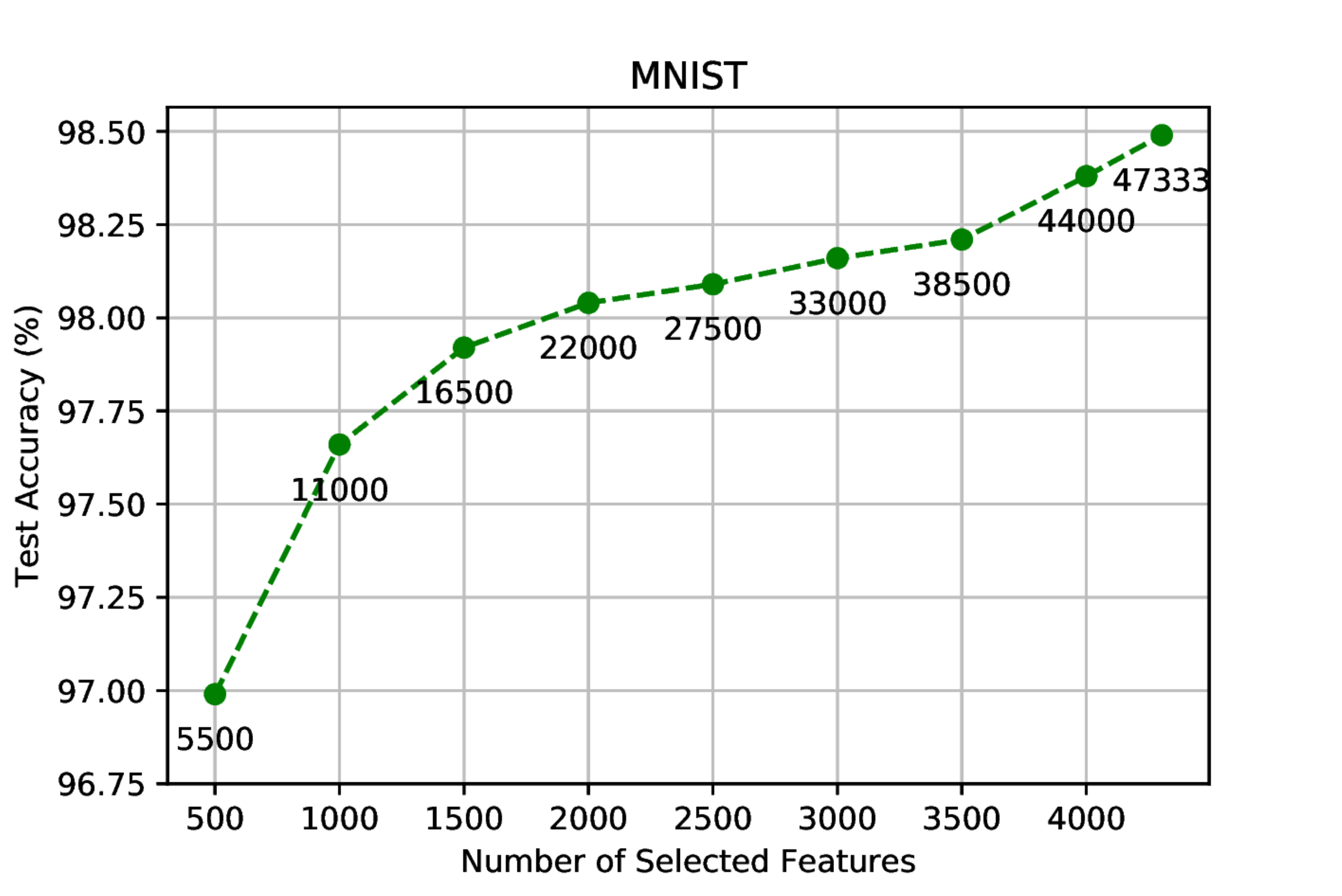}}
\end{minipage} 
\hfill
\begin{minipage}[b]{0.31\linewidth}
\centering
\centerline{\includegraphics[width=0.97\linewidth]{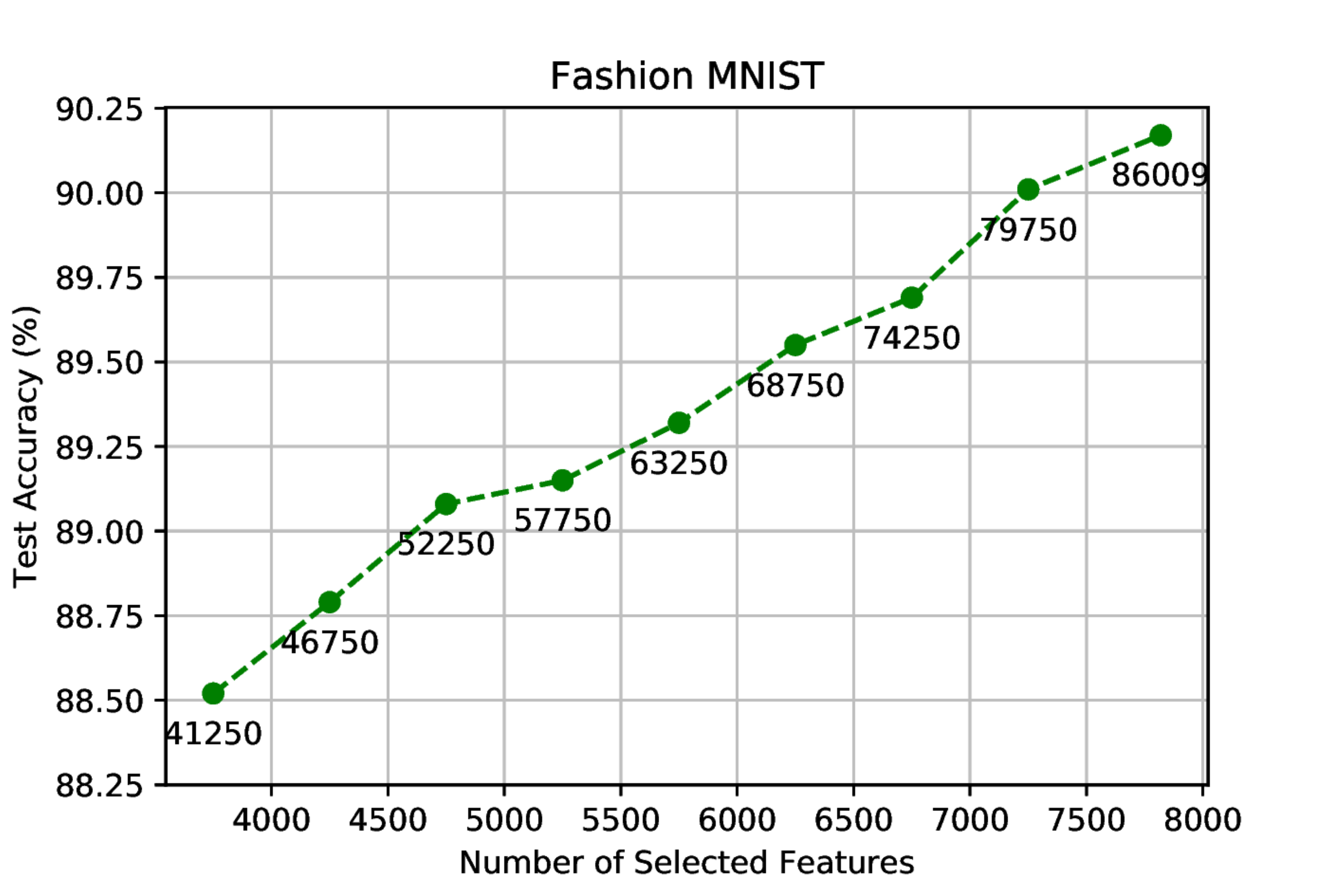}}
\end{minipage}
\hfill
\begin{minipage}[b]{0.31\linewidth}
\centering
\centerline{\includegraphics[width=0.97\linewidth]{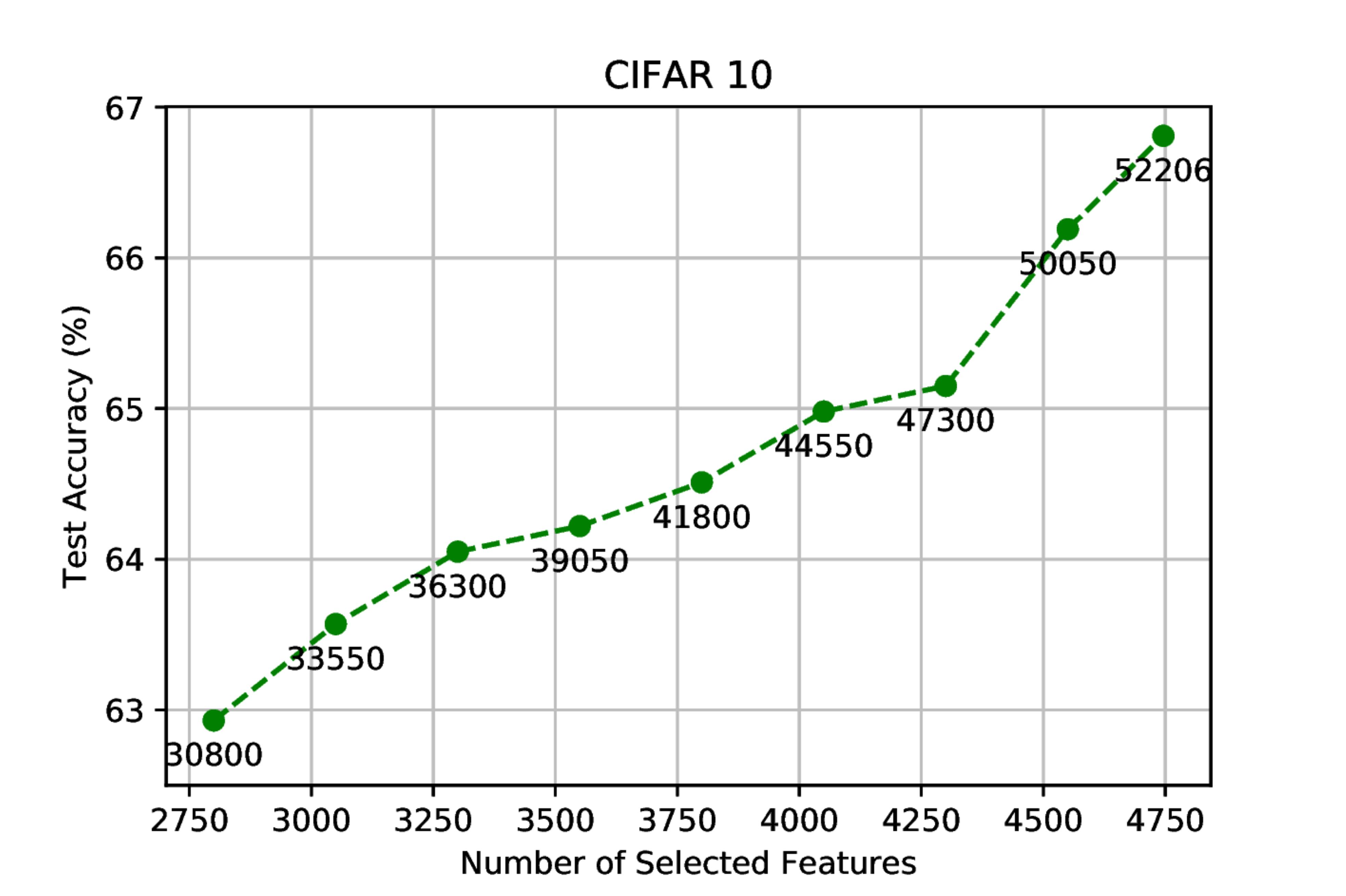}}
\end{minipage}
\caption{The relation between test accuracy (\%) and selected number
$N_S$ of cross-entropy-guided features in PixelHop++ for MNIST, Fashion
MNIST and CIFAR-10, where the number of model parameters in Module 2 is
shown at each operational point.} \label{fig:cross}
\end{figure*}


\subsection{Experimental Setup}

We test the PixelHop++ method on three popular datasets: MNIST
\cite{Lecun98gradient-basedlearning}, Fashion MNIST
\cite{xiao2017fashion} and CIFAR-10 \cite{krizhevsky2009learning}.
MNIST and Fashion MNIST contain gray-scale images of size $28 \times 28$
and zero-padding is used to enlarge the image size to $32\times32$.
CIFAR-10 has 10 object classes of color images of size $32\times32$. We
conduct performance benchmarking of PixelHop++ and LeNet-5
\cite{Lecun98gradient-basedlearning} in terms of classification accuracy
and model complexity. We adopt the original LeNet-5 architecture for
MNIST. To handle more complicated images in Fashion MNIST and CIFAR-10,
we increase the filter numbers of the convolutional layers and
fully-connected layers as shown in Table \ref{table:architecture}.  By
applying PixelHop++ to MNIST, we adopt the tree-decomposed feature
representation and use features of leaf nodes from the third level only
since the discriminant power of features in the first two levels is very
weak.  For Fashion MNIST and CIFAR-10, we use concatenated features from
leaf nodes at all three levels.  We set the output dimension, $M$, of
the LAG unit to $M=10$. 

\begin{table}[!htb]
\begin{center}
\footnotesize
\caption{Comparison of the original and the modified LeNet-5
architectures on three benchmark dataset.}\label{table:architecture}
\begin{tabular}{cccc} \hline
Dataset  & MNIST  & Fashion MNIST & CIFAR-10 \\ \hline
1st Conv. Kernel Size & $5 \times 5 \times 1$ & $5 \times 5 \times 1$ &  $5 \times 5 \times 3$ \\ \hline
1st Conv. Kernel No.  & $6$ &$16$ &  $32$     \\ \hline
2nd Conv. Kernel Size & $5 \times 5 \times 6$ &$5 \times 5 \times 16$  &  $5 \times 5 \times 32$ \\ \hline
2nd Conv. Kernel No. & $16$ &$32$ & $64$  \\ \hline
1st FC. Filter No. & $120$& $200$ & $200$ \\ \hline
2nd FC. Filter No. & $84$ & $100$ & $100$ \\ \hline
Output Node No.& $10$ & $10$ & $10$  \\ \hline
\end{tabular}
\end{center}
\end{table}

\subsection{Effects of Hyper-Parameters in PixelHop++}\label{subsec:hyper}

We study the effect of two hyperparameters, $T$ and $N_S$, on
classification accuracy of PixelHop++. We can control the model size in
Module 1 by energy threshold $T$. A larger threshold demands a larger
model size. As shown in Fig. \ref{fig:energy}, the test accuracy
decreases slightly as $T$ decreases while the model size is reduced
significantly for all three datasets. For example, by moving $T$ from
0.00001 and 0.0005 for MNIST, the test accuracy is decreased by
0.51\% while the model parameter number becomes 4x fewer. We see clear
advantages of the tree-decomposed feature representation in finding good
tradeoff. 

By varying the $N_S$ values in Module 2, we plot the test accuracy as a
function of $N_S$ in Fig.  \ref{fig:cross}. The test accuracy decreases
marginally as we decrease $N_S$ to yield smaller models for all three
datasets. Take MNIST as an example, the test accuracy of keeping
500 features is only 0.57\% lower than keeping all features (3x more) in Module 2. This is because a smaller feature subset with more
discriminant features can be selected through the cross-entropy-guided
feature selection process.  Overall, by adjusting $T$ and $N_s$, we can
control the model size in fine-granularity. 

\subsection{Performance Benchmarking}\label{subsec:complexity}

We compare classification accuracy and model complexity of LeNet-5 and
PixelHop++ against all three datasets in Table \ref{table:accuracy_1}
and Table \ref{table:parameters}, respectively. Here, we report the
performance of two model settings of PixelHop++, {\em i.e.} a larger
model and a smaller model.  In terms of classification accuracy, LeNet-5
performs the best on MNIST and CIFAR-10, yet the large PixelHop++ model outperforms
LeNet-5 on Fashion MNIST with fewer parameters.  On the
other hand, by slightly scarifying the accuracy, PixelHop++ with the
small model demands about 2x, 6x and 6x fewer parameters than LeNet-5
for MNIST, Fashion MNIST and CIFAR-10, respectively. With these
performance numbers, we can claim that PixelHop++ is more effective than
LeNet-5. 

\begin{table}[h!]
\centering
\small
\caption{Comparison of test accuracy (\%) of LeNet-5 and
PixelHop++ for MNIST, Fashion MNIST and CIFAR-10. }\label{table:accuracy_1}
\begin{tabular}{cccc} \hline
 Method    & MNIST  & Fashion MNIST & CIFAR-10 \\ \hline
 LeNet-5   & 99.04 & 89.74 & 68.72 \\ \hline
 PixelHop++ (Large)  &98.49 & 90.17 & 66.81  \\\hline
 PixelHop++ (Small)  &97.98 & 88.84 & 64.75  \\\hline
\end{tabular}
\end{table}


\begin{table}[h!]
\centering
\small
\caption{Comparison of the model size (in terms of the total parameter
numbers) of LeNet-5 and PixelHop++ for the MNIST, the Fashion
MNIST and the CIFAR-10 datasets.} \label{table:parameters}
\begin{tabular}{cccc} \hline
 Method& MNIST  & Fashion MNIST & CIFAR-10 \\ \hline
 LeNet-5 & 61,706 & 194,558 & 395,006 \\ \hline
 PixelHop++ (Large) &  111,981 & 127,186 & 115,623 \\ \hline
 PixelHop++ (Small) &  29,514 & 33,017 & 62,150\\ \hline
\end{tabular} \\
\end{table}

\section{Conclusion}\label{sec:conclusion}

An image classification method with a design of interpretable and small
learning models was proposed in this paper. Extensive experiments were
conducted on three benchmark datasets (MNIST, Fashion MNIST, and
CIFAR-10) to demonstrate that the model size of PixelHop++ can be
flexibly controlled, and PixelHop++ maintain the classification accuracy
with fewer parameters comparing with LeNet-5.

\newpage
\bibliographystyle{IEEEbib}
\bibliography{refs}

\end{document}